# Discriminating BCC Subtypes Using Entropy and Mutual Information from Dermoscopic Features

Iván Matas[1], Begoña Acha[1], Francisca Silva-Clavería[2], Amalia Serrano[2], Tomás Toledo-Pastrana[3], and Carmen Serrano[1]

[1] Department of Signal Theory and Communications, Higher Technical School of Engineering, University of Seville, Seville, Spain, [2]Dermatology Service, Virgen Macarena University Hospital, Seville, Spain, and [3]Quironsalud Infanta Luisa and Sagrado Corazón Hospitals, Seville, Spain

**Abstract**

**Objective:** To analyze the frequency of occurrence of different dermoscopic patterns in BCC lesions, the co-occurrence of pattern pairs, and their relationship with the histopathologic subtype of BCC, using statistical methods and tools from Information Theory such as entropy, conditional entropy, mutual information, and Hamming weight.

**Methods:** A total of 223 dermoscopic images (256×256 pixels) of histologically confirmed BCC lesions from Hospital Universitario Virgen Macarena (Seville, Spain) were analyzed. Each image was multilabel-annotated for the presence of nine dermoscopic patterns and categorized into one of four BCC subtypes: superficial, nodular, infiltrative, or micronodular. A statistical and information-theoretic framework was applied, including co-occurrence matrices, Bayesian conditional probabilities, and entropy-based measures. Mutual information was used to quantify the predictive value of individual patterns and pattern combinations, and decision trees were constructed based on their diagnostic informativeness.

**Results:** Nodular BCC showed high association with most dermoscopic patterns, especially arborizing telangiectasia and blue-gray ovoid nests. In contrast, superficial BCC was more frequently linked to maple leaf-like structures and shiny white-red areas. Certain patterns such as spoke-wheel areas and white streaks showed low discriminative power. Mutual information and conditional probabilities revealed relevant pattern-pair dependencies for each subtype. The decision trees demonstrated how subsets of patterns could optimize subtype classification through the accumulation of diagnostic information.

**Conclusions:** The integration of Information Theory provides a quantitative and nuanced understanding of the relationships between dermoscopic patterns and BCC subtypes. This approach highlights key diagnostic patterns, supports the differentiation of complex cases, and establishes a foundation for the development of automated, pattern-based diagnostic tools. Further studies with larger and more balanced datasets are recommended to generalize and strengthen these findings.

## 1. Introduction

Basal Cell Carcinoma (BCC) is the most common form of skin cancer, with approximately 3.6 million cases diagnosed in the United States each year [1]. In order to detect them early – so the treatment and cure could be easier – understanding risk factors, BCC causes and warning signs can help.

In addition to clinical diagnosis, dermatoscopy has also been found to be a useful tool in the preoperative prediction of the BCC subtype and in the non-invasive assessment of tumor response to topical treatments. However, sometimes, the BCC subtype has to be assessed histopathologically [2].

Dermatoscopic criteria for BCC include absence of pigment network, arborizing and superficial telangiectasia, multiple erosions, ulceration, bluish-grey clods of variable size (ovoid nests and globules and focused dots), leaf-like areas, spoke-wheel areas, and clods within a clod (concentric structure) [3].

**The treatment of BCC depends on its histopathologic subtype, and to avoid unnecessary biopsies, various studies have sought to identify BCC subtypes through dermoscopy.**

Lallas et al. [4] aimed to differentiate superficial BCC (sBCC) from other subtypes by analyzing 77 sBCCs and 258 non-sBCCs, concluding that maple leaf-like areas, superficial telangiectasia, multiple small erosions, and shiny white-red structureless areas were strong predictors of sBCC, while arborizing vessels, blue-gray ovoid nests, and ulceration were more indicative of non-sBCCs. Their model achieved a sensitivity of 81.9% and a specificity of 81.6%. Similarly, Popadic et al. [5] evaluated the accuracy of dermoscopy in diagnosing subtypes, analyzing 102 histopathologically verified BCCs (superficial, nodular, and aggressive). They found the best performance in superficial BCCs, with good correlation for nodular types and lower accuracy for aggressive types. Longo et al. [6] expanded on these efforts by identifying dermatoscopic patterns in 22 superficial, nodular, and infiltrative BCCs. Superficial fine telangiectasia, shiny white-red structureless areas, and multiple small erosions were prevalent in sBCCs, while nodular BCCs displayed arborizing vessels and blue-gray ovoid nests, and infiltrative types were associated with shiny white-red structureless areas and arborizing vessels. Together, these studies emphasize the predictive value of specific dermatoscopic features, particularly for sBCCs, but highlight variability in diagnostic performance for more aggressive subtypes.

Further exploring vascular patterns, Lupu et al. [7] reviewed dermoscopic features in up to 609 BCC lesions across various subtypes. They noted that arborizing vessels were strongly associated with nodular BCC (57.1%-82% prevalence in pigmented BCCs), while short fine telangiectasias were highly specific to non-pigmented sBCCs (92%). Milky-pink backgrounds were common in sBCCs but lacked diagnostic specificity, and aggressive subtypes displayed unique vessel morphologies like glomerular and truncated vessels. Similarly, Pogorzelska-Dyrbuś et al. [8] analyzed 120 BCC lesions across different subtypes and anatomical zones. Arborizing and short fine telangiectasias were the most frequent patterns, with superficial BCCs in non-H zones showing a higher prevalence of comma and glomerular vessels, potentially reflecting more superficial tumor characteristics in these areas.

Finally, broader reviews and systematic analyses have sought to integrate findings across subtypes. Reiter et al. [9] summarized 31 studies comprising 5,950 BCC cases, identifying arborizing vessels as the most common pattern (59%), particularly in nodular and infiltrative subtypes. Shiny white structures were frequent across all subtypes (49%), while blue-grey ovoid nests were predominant in pigmented and nodular BCCs. Emiroglu et al. [10], examining 98 BCC cases, found arborizing vessels in 42.9% of cases, with thicker vessels (>0.2 mm) in nodular BCCs and finer vessels in superficial types. They also identified patterns like milky-pink backgrounds in superficial BCCs and blue-gray globules in pigmented subtypes. Popadić et al. [11], analyzing 151 lesions, reported distinct patterns for each subtype: nodular BCCs were characterized by a milky red background and translucency, while superficial BCCs frequently showed multiple erosions and hypopigmentation. Pigmented BCCs displayed blue-gray ovoid nests, and aggressive subtypes like morpheaform, and infiltrative BCCs were associated with arborizing microvessels and ulceration.

Despite the significant body of research on dermoscopic patterns across different BCC subtypes, certain gaps remain unaddressed. To our knowledge, no studies have applied advanced concepts from information theory, such as entropy, conditional entropy, and mutual information, for in-depth analysis of the variability, uncertainty, and dependencies among dermoscopic patterns in BCC images. Entropy provides a quantitative measure of variability in the distribution of patterns, while conditional entropy evaluates the remaining uncertainty in one pattern given the presence of another. Mutual information quantifies the statistical dependency between co-occurring patterns, offering insights into their diagnostic relevance. Additionally, the level of consensus among specialists regarding the identification of specific patterns and the agreement between AI-based tools and dermatologists in recognizing these patterns have not been systematically assessed. Such evaluations are crucial for validating the consistency and diagnostic accuracy of dermoscopic criteria across various BCC subtypes, as reported in existing literature.

Furthermore, Information Theory applied to analyze co-occurrence of BCC pattern pairs may provide a deeper understanding of the complexity and diagnostic significance of dermoscopic findings.This study aims to expand the understanding of BCC patterns beyond traditional approaches, incorporating quantitative methods to address variability and strengthen diagnostic criteria.

## 2. Methods

### 2.1 Database elaboration

The database contains 223 dermoscopic images (256×256 pixels) from the Hospital Universitario Virgen Macarena of Seville, Spain. Each image can present one or more BCC patterns, making it a multilabel classification problem. The patterns include pigment network (PN), multiple blue-gray globules (MG), large blue-gray ovoid nests (BO), maple leaf-like (ML), arborizing telangiectasia (AT), ulceration (UL), spoke wheel (SW), shiny white-red structure (SWRS), and white streaks (WS). Additionally, biopsy results categorize the images into four BCC subtypes: Infiltrative, Micronodular, Nodular, and Superficial. Table 1 summarizes the distribution of patterns across subtypes, highlighting the complexity and diversity of the dataset.

Patterns present in each image have been labelled by one dermatologist, codifying the presence or absence of each pattern for a specific image with a 1/0 value respectively, in a binary word of size 1×9.

This study was performed in line with the principles of the Declaration of Helsinki. Approval was granted by the Ethical Committee for Biomedical Research from Andalusia (protocol code: 1901-N-22).

### 2.2 Statistical analysis

#### 2.2.1 Statistical analysis

We conducted a detailed statistical analysis of the patterns present in the BCC database to better understand their relationships and interactions across different BCC subtypes. This analysis was structured as follows:

- **Co-occurrence of Patterns:**

We developed a co-occurrence matrix to quantify how often dermoscopic patterns and BCC subtypes appear together. This matrix is designed to systematically capture the frequency of these pairings, providing a structured way to explore their associations. By examining both common and less frequent combinations, the analysis offers a nuanced view of the underlying relationships. This approach not only highlights clear patterns but also uncovers subtle connections that might otherwise be overlooked. Such insights are instrumental in enhancing diagnostic accuracy, as they help clinicians better understand how specific dermoscopic features correspond to BCC subtypes. Through this method, the interplay between dermoscopic features and BCC subtypes becomes clearer, supporting more informed decision-making in both research and practice.

This calculation helps to identify patterns that are characteristic of specific subtypes and provides insights into potential diagnostic indicator, supporting the differentiation of BCC subtypes based on their unique dermoscopic features.

- **Bayesian Probability Analysis:**

To deepen our understanding of the relationship between patterns and subtypes, we performed a Bayesian analysis to extract relevant conditional probabilities. Specifically, we calculated:

1. $P(class|pattern)$: The probability of a BCC subtype given the presence of a particular pattern, providing insight into the likelihood that a subtype is associated with each dermoscopic characteristic.

2. $P(pattern|class)$: The probability of observing a specific pattern given a BCC subtype, revealing which patterns are most characteristic of each subtype.

3. Furthermore, we extended this analysis to consider conditional probabilities involving combinations of patterns, such as $P(class|PatternA,PatternB)$, to examine how pairs of patterns might jointly influence the probability of a given BCC subtype. This approach aims to capture nuanced interactions between patterns that single-pattern probabilities may miss, providing a more complex understanding of how combinations of features can indicate specific BCC types.

**2.3 Information Theory**

Information theory is a mathematical discipline that studies the quantification, transmission, and efficient storage of data. Introduced by Claude Shannon in 1948 [12], it has become a cornerstone of modern telecommunications. Its significance lies in enabling the design of transmission systems capable of handling large volumes of data, minimizing information loss, and optimizing the use of the electromagnetic spectrum. These capabilities are essential for contemporary technologies such as mobile networks, the internet, and satellite communications, where the efficient and reliable transfer of information is crucial. Key concepts such as entropy and mutual information are instrumental in analysing and improving the performance of coding, compression, and transmission systems, ensuring that data reaches its destination effectively with minimal loss or error.

Furthermore, information theory extends its applicability beyond telecommunications and can be employed to study relationships between variables in clinical research. By quantifying uncertainty through entropy and statistical dependency through mutual information, it provides a robust framework for analyzing complex interactions. For instance, in a clinical context where one variable represents a subtype of BCC and another denotes such BCC patterns, mutual information can reveal how much knowledge of the BCC patterns reduces uncertainty about the diagnostic. This ability to detect intricate, potentially non-linear dependencies makes information theory a powerful tool for uncovering insights in medical data and enhancing the understanding of clinical phenomena.

**2.3.1 Entropy**

Entropy, denoted by $H$, is a central concept in information theory that measures the uncertainty associated with a random variable. Mathematically, it is defined as:

$$H(X) = -\sum_{x \in X} P(x) \log_2 P(x) \tag{1}$$

where $X$ is a random variable, $x$ represents its set of possible values, and $P(x)$ is the probability of each value. Intuitively, entropy quantifies the average amount of information needed to describe a randomly chosen event from a data source. In a clinical context, entropy can be used to evaluate the unpredictability of a diagnostic indicator. High entropy indicates greater variability or uncertainty in the data, often requiring more sophisticated modelling or analysis to capture the underlying patterns.

**2.3.2 Conditional Entropy $H(\text{Class} \mid \text{Pattern})$**

Conditional entropy quantifies the uncertainty about the classes given that the patterns are known. It provides insight into how much uncertainty remains in classifying an observation when specific BCC patterns are observed. Mathematically, it is defined as:

$$H(Y \mid X) = \sum_{x \in X} P(x) \cdot \left[ -\sum_{y \in Y} P(y \mid x) \log_2 P(y \mid x) \right] \tag{2}$$

In the context of dermoscopic analysis for BCC subtypes, this measure helps to evaluate the diagnostic value of specific patterns by determining how much ambiguity remains in assigning a subtype based on observed features. A low conditional entropy indicates that the pattern strongly predicts the class, whereas a high conditional entropy suggests significant overlap or ambiguity in classification.

**2.3.3 Mutual Information**

Mutual information, denoted as $I(X;Y)$, measures the statistical dependency between two random variables $X$ and $Y$. It is defined as:

$$I(X;Y) = \sum_{x \in X} \sum_{y \in Y} P(x,y) \log_2 \frac{P(x,y)}{P(x)P(y)} \tag{3}$$

This measure quantifies how much knowing one variable reduces the uncertainty about the other. In clinical data analysis, mutual information can be instrumental in understanding the relationship between two variables, such as a BCC subtype $X$ and a BCC pattern $Y$. For example, if $I(X;Y)$ is high, the BCC patterns provide significant information about the BCC subtype, indicating its potential as a predictor.

Alternatively, mutual information can be expressed using entropy and conditional entropy as:

$$I(X;Y) = H(X) - H(X \mid Y) = H(Y) - H(Y \mid X) \tag{4}$$

This formulation shows that mutual information quantifies how much knowing one variable reduces the uncertainty about the other. In clinical data analysis, mutual information can be instrumental in understanding the relationship between two variables, such as a BCC subtype $X$ and a BCC pattern $Y$. For example, if $I(X;Y)$ is high, indicates that the pattern provides substantial predictive information about the subtype, linking uncertainty (entropy) with dependency.

**2.3.4 Hamming Weight**

The Hamming weight is a concept from coding theory that counts the number of non-zero elements in a binary string. Formally, given a binary string $b_i$, the Hamming weight, denoted as $\overline{W}_H$, is defined as:

$$\overline{W}_H = \frac{1}{N} \sum_{i=1}^{N} b_i \tag{5}$$

where $b_i$ represents the *i-th* bit of the binary string $b$, and *n* is the total number of bits in the string. In the context of information theory, the Hamming weight is often used to evaluate the efficiency or error-detecting capability of a code by measuring the minimum distance between codewords.

In clinical data analysis, the concept of Hamming weight can be applied to binary encoded features derived from medical images or diagnostic indicators. For example, when BCC patterns are encoded as binary vectors, the Hamming weight can represent the number of positive detections of patterns in a specific BCC subtype. By analysing the distribution of Hamming weights in a dataset, researchers can assess the complexity or sparsity of the patterns and their possible diagnostic relevance for certain types of BCC. In other words, it is not the same information if 8 patterns are activated in 2 samples as in 10.

This statistical approach is essential to validate the correlation between dermoscopic patterns and their frequencies across BCC subtypes reported in previous studies. Additionally, by exploring pattern co-occurrence, Bayesian probabilities, and mutual information, we aim to enhance the reliability of dermatoscopic criteria in differentiating BCC subtypes. The entropy analysis provides a quantitative measure of variability and uncertainty in pattern distributions, while mutual information quantifies the dependency between co-occurring patterns.

**3. Results**

**3.1 Patterns as indicators of BCC Subtype**

Upon examining Figure 1.1 which represents $P(\text{class} | \text{pattern})$, several notable findings emerge in the characterization of BCC subtypes based on observed dermoscopic patterns. First, the nodular subtype consistently displays a high probability, specifically greater than 0.6, across a majority of patterns, with 66% of the patterns meeting this threshold. When excluding the PN pattern, which serves as a negative criterion, 75% of the remaining patterns exhibit probabilities exceeding 0.6. This observation suggests that the presence of any pattern other than SW and SWRS strongly supports a diagnosis of the nodular subtype of BCC. Consequently, if a BCC pattern —excluding SW and SWRS— is identified, there is strong evidence to suggest that the subtype in question is nodular BCC. This makes the nodular subtype the most distinguishable among the BCC subtypes in this analysis, as it is associated with high probabilities across most BCC patterns.

In contrast, the SW pattern shows a relatively uniform distribution across all BCC subtypes, indicating that it provides limited diagnostic information for subtype classification. This consistency across subtypes implies that the SW pattern has low entropy, making it less informative in distinguishing between different BCC subtypes. Therefore, the presence of the SW pattern alone does not substantially aid in determining the specific subtype, as it appears relatively equally among them. Within the context of this analysis, SW emerges as the least informative pattern due to its limited discriminative power.

The ML pattern appears to be relevant for the superficial subtype. Excluding the nodular subtype, the probability of ML is significantly higher for the superficial subtype than for the other two deeper subtypes. On the other hand, BO may be indicative of the infiltrative subtype, as it demonstrates a higher probability for this specific subtype. These findings highlight the nuanced diagnostic value of certain dermoscopic patterns in differentiating between BCC subtypes.

Further, the ML and SWRS patterns exhibit higher entropy compared to other patterns, as they are present across multiple subtypes. Specifically, the ML pattern is predominantly observed in the nodular and superficial subtypes, while SWRS appears across infiltrative, nodular, and superficial subtypes. This broader distribution suggests that these patterns lack specificity for a single BCC subtype, making them more ambiguous in their diagnostic implications. Notably, if the nodular subtype is excluded —given that most patterns appear in this category— the presence of ML or SWRS may serve as a useful indicator for identifying either infiltrative or superficial BCC. This observation highlights the potential diagnostic value of ML and SWRS patterns in differentiating infiltrative and superficial BCC when the nodular subtype is less likely.

Table V details the Hamming weight analysis described in Section 2.3.4, highlighting that the Infiltrative subtype is associated with a markedly lower presence of dermoscopic patterns, whereas the Superficial subtype exhibits the highest prevalence. Lesions characterized by numerous patterns are likely indicative of the Superficial subtype, while those with sparse patterns may correspond to the Infiltrative subtype.

### 3.2 Distribution of BCC Patterns Across Subtypes

In the Figure 1.2 which represents $P(\text{pattern} \mid \text{class})$, several important observations emerge regarding the distribution of BCC patterns across different BCC subtypes. Patterns such as AT and UL show significant presence across all BCC subtypes, although certain variations are noted among specific subtypes. The widespread presence of AT and UL suggests that these patterns are general indicators of BCC but offer limited specificity in distinguishing between subtypes.

Notably, the MG pattern is highly prevalent in both micronodular and superficial subtypes, indicating that it may serve as a distinguishing feature for these subtypes. This prevalence suggests that when MG is observed, there is a higher likelihood that the BCC subtype is either micronodular or superficial. Furthermore, the ML pattern is prominently present in the superficial subtype, while its presence in other subtypes is considerably lower. This distinctive distribution makes ML a potential marker for identifying the superficial subtype, as its probability is notably elevated in this subtype compared to others.

The SWRS pattern appears with high probability in the superficial subtype and, to a lesser extent, in the infiltrative subtype. This distribution implies that SWRS might have diagnostic relevance when differentiating between superficial and infiltrative subtypes, though its association is strongest with superficial BCC. Such a pattern could aid in cases where superficial BCC is suspected, especially in the absence of stronger indicators for other subtypes.

Finally, the WS and SW patterns show minimal diagnostic utility, as they have relatively low probabilities across all subtypes. Their low prevalence suggests that these patterns contribute limited information for identifying specific BCC subtypes, making them less reliable features in clinical assessments.

### 3.3 Can we establish some relationship between patterns for each subtype of BCC?

Figure 2 illustrates the co-occurrence matrices of patterns for each BCC subtype, presented in absolute values without normalization for the total number of samples or subtype sample sizes. These results align with findings reported in the literature to date. For infiltrative BCC, high correlations are observed between the pairs MG-ML, MG-AT, SWRS-ML, and SWRS-MG, with a notable prevalence of the AT, UL, MG, and SWRS patterns, identifying AT, UL, and SWRS as key determinants for detecting this subtype. Micronodular BCC shows strong correlations between the UL-AT and AT-MG patterns, consistent with results from the likelihood analysis. Nodular BCC exhibits correlations in the pairs AT-MG, BO-AT, BO-UL, UL-SWRS, and AT-UL. Finally, superficial BCC displays high correlations among the patterns MG-ML, MG-AT, MG-SWRS, and ML-SWRS.

### 3.4 Can we establish some relationship between two patterns for each subtype of BCC?

As outlined in Section 2.3, the innovation of this paper lies in the application of information theory from telecommunications. To achieve this, we integrate the conditional probabilities $P(class \mid PatternA, PatternB)$ presented in Table VI. A threshold of 0.7 was established for improved visualization, as lower probabilities contribute less information and result in a table of excessive length. Additionally, using Equation 4, we calculated the mutual information $I(class; pattern)$ and $I(pattern; class)$, which are shown in Figures 3.1 and 3.2, respectively. Of these, $I(class; pattern)$ is of particular importance, as the presence of patterns provides insight into the BCC subtype being analysed.

To summarize this information, a decision tree was constructed by combining the data from Figure 3.1 with Table 4. For each BCC subtype, the decision tree begins with the pattern that provides the highest information. Nodes are subsequently added based on the joint probabilities in Table 4, representing relationships among patterns. Each node contains the following details: the acronym of the corresponding pattern, a binary value indicating the presence (1) or absence (0) of the pattern, and the accumulated mutual information, which is the sum of the mutual information provided by the *i-th* pattern and the previously accumulated mutual information.

To simplify the graph, if two nodes represent the same pattern with the same value, only the node with the highest mutual information in its upper section is extended. This ensures that the final graph prioritizes pathways that convey the most information.

### 4. Conclusion

This research leveraged advanced telecommunication methodologies to comprehensively analyse the association between patterns and BCC subtypes. By employing a systematic approach, it revealed crucial insights with significant implications for both clinical practice and dermatological research.

The analysis established that the nodular subtype exhibits a strong correlation with most BCC patterns, excluding SW and SWRS, underscoring its distinct diagnostic features. Notably, patterns such as ML and SWRS were identified as having elevated entropy and broader distributions, demonstrating their utility in differentiating infiltrative and superficial subtypes. Conversely, the SW pattern was found to have limited diagnostic specificity, further emphasizing the need for its cautious interpretation.

The exploration of co-occurrence relationships among patterns elucidated key interdependencies, such as the association between MG and ML patterns in superficial BCC and significant mutual information for AT and UL patterns in infiltrative BCC. These findings, integrated into a decision tree model, enhance the precision of subtype classification and reinforce the importance of pattern-based diagnostics.

However, the study's findings are tempered by limitations, primarily the unequal distribution of samples across subtypes, with a marked underrepresentation of micronodular and superficial cases. This imbalance may restrict the generalizability of the results and highlights the necessity of future studies incorporating larger and more diverse datasets.

Future work should focus on leveraging machine learning algorithms to further refine the diagnostic potential of BCC patterns. By integrating computational techniques with conditional probabilities and mutual information, an automated diagnostic framework could be developed for real-time clinical application. Additionally, extending the dataset to encompass a more representative distribution of subtypes and populations will enhance the robustness and applicability of the findings.

In summary, this study advances the field by offering a detailed understanding of BCC patterns in BCC subtypes, providing a solid foundation for future research and the development of diagnostic tools that can improve clinical diagnosis in dermatology.

## 5. Discussion

The characterization of BCC subtypes is integral to enhancing diagnostic accuracy and optimizing clinical decision-making. This study provides a robust framework for understanding the diagnostic value of specific dermoscopic patterns and their role in distinguishing among BCC subtypes.

Key findings highlight that AT and BO patterns are the most reliable indicator for nodular BCC, offering critical diagnostic clarity. However, while SWRS is frequently observed in nodular cases, its limited specificity due to cross-subtype occurrence underscores the need for contextual and integrative diagnostic strategies.

For superficial BCC, short fine telangiectasia and ML patterns emerged as significant indicators. Although ML patterns overlap with nodular cases, their relevance for superficial subtypes is underscored when combined with clinical observations. The BO pattern—predominantly linked to infiltrative BCC—was validated as a highly reliable sign, emphasizing its diagnostic utility in challenging cases.

Conversely, the SW pattern, consistently observed across all subtypes, demonstrated limited discriminatory capacity. Despite its broad presence, its diagnostic value remains ancillary. Patterns such as ML and SWRS, characterized by broader distributions, hold potential as supplementary indicators for differentiating superficial and infiltrative BCC in complex scenarios.

The study's limitations, including the unequal distribution of samples among subtypes, particularly micronodular and superficial cases, must be acknowledged. This imbalance necessitates future research involving larger and more representative datasets to validate and extend the current findings.

Looking ahead, the application of machine learning techniques presents an opportunity to enhance the diagnostic potential of dermoscopic analysis. By employing advanced computational tools, future research can develop automated frameworks that integrate the insights gained from this study into real-time clinical workflows. Moreover, diversifying the dataset to encompass broader populations and BCC subtypes will strengthen the generalizability of the results.

**TABLE LEGENDS**

**Table I. Example of multilabel and binary encoding for BCC diagnosis**

| Codification | Multilabel | Binary | Diagnosis |
|---|---|---|---|
| Example 1 | [0 1 0 1 1 0 1] | 1 | BCC Presence |
| Example 2 | [1 0 0 0 0 0 0] | 0 | BCC Absence |
| Example 3 | [0 0 0 0 0 0 0] | 0 | BCC Absence |

**Table II. Database distribution**

| Subtype BCC | Samples | BCC patterns | | | | | | | | |
|---|---|---|---|---|---|---|---|---|---|---|
| | | PN | MG | BO | ML | AT | UL | SW | SWRS | WS |
| Infiltrative | 42 | 0 | 8 | 9 | 4 | 22 | 20 | 2 | 12 | 3 |
| Micronodular | 15 | 0 | 6 | 3 | 2 | 10 | 7 | 2 | 2 | 0 |
| Nodular | 142 | 0 | 42 | 39 | 27 | 78 | 79 | 4 | 31 | 11 |
| Superficial | 24 | 0 | 10 | 2 | 11 | 11 | 8 | 3 | 11 | 12 |
| Total | 223 | 0 | 66 | 53 | 44 | 121 | 114 | 11 | 56 | 26 |

**Table III. Frequency distribution**

| Subtype BCC | P(Class) | $P(pattern \mid \text{class})$ | | | | | | | | |
|---|---|---|---|---|---|---|---|---|---|---|
| | | PN | MG | BO | ML | AT | UL | SW | SWRS | WS |
| Infiltrative | 0.19 | 0 | 0.19 | 0.21 | 0.10 | 0.52 | 0.47 | 0.05 | 0.29 | 0.07 |
| Micronodular | 0.07 | 0 | 0.40 | 0.20 | 0.13 | 0.66 | 0.46 | 0.13 | 0.13 | 0 |
| Nodular | 0.63 | 0 | 0.29 | 0.27 | 0.19 | 0.55 | 0.55 | 0.03 | 0.21 | 0.08 |
| Superficial | 0.10 | 0 | 0.42 | 0.08 | 0.45 | 0.46 | 0.33 | 0.12 | 0.46 | 0.08 |
| P(Pattern) | | 0 | 0.29 | 0.23 | 0.19 | 0.54 | 0.51 | 0.05 | 0.25 | 0.11 |

**Table IV. Frequency distribution**

| Patterns | P(Pattern) | $P(\text{class} \mid pattern)$ | | | |
|---|---|---|---|---|---|
| | | Infiltrative | Micronodular | Nodular | Superficial |
| PN | 0 | 0.00 | 0.00 | 0.00 | 0.00 |
| MG | 0.29 | 0.12 | 0.09 | 0.64 | 0.15 |
| BO | 0.23 | 0.17 | 0.06 | 0.74 | 0.04 |
| ML | 0.19 | 0.09 | 0.05 | 0.61 | 0.25 |
| AT | 0.54 | 0.18 | 0.08 | 0.64 | 0.09 |
| UL | 0.51 | 0.18 | 0.06 | 0.69 | 0.07 |
| SW | 0.05 | 0.18 | 0.18 | 0.36 | 0.27 |
| SWRS | 0.25 | 0.21 | 0.04 | 0.55 | 0.20 |
| WS | 0.11 | 0.19 | 0.00 | 0.69 | 0.12 |
| P(Class) | | 0.19 | 0.07 | 0.63 | 0.10 |

**Table V. Hamming weight**

| Subtype BCC | Infiltrative | Micronodular | Nodular | Superficial |
|---|---|---|---|---|
| $\overline{W}$ | 1,90 | 2,13 | 2,19 | 2,41 |

**Table VI. Conditional probabilities for BCC subtype predictions based on paired dermoscopic patterns. The table presents P( class | PatternA,PatternB )values above the 0.7 threshold were filtered to highlight significant associations, highlighting how specific pattern combinations jointly influence subtype probability.**

Pigment network (PN), multiple blue-gray globules (MG), large blue-gray ovoid nests (BO), maple leaf like (ML), arborizing telangiectasia (AT), ulceration (UL), spoke wheel (SW), shiny white-red structure (SWRS), white streaks (WS).

| P( class \| PatternA,PatternB ) | Pattern A | Pattern B | Value A | Value B |
|---|---|---|---|---|
| **Infiltrative BCC** | | | | |
| 0.99 | UL | SW | 1 | 1 |
| **Nodular BCC** | | | | |
| 0.70 | BO | AT | 1 | 1 |
| 0.71 | MG | UL | 0 | 1 |
| 0.71 | UL | SWRS | 1 | 0 |
| 0.71 | AT | UL | 1 | 1 |
| 0.71 | UL | WS | 1 | 1 |
| 0.73 | BO | SWRS | 1 | 0 |
| 0.73 | BO | WS | 0 | 1 |
| 0.74 | PN | BO | 0 | 1 |
| 0.75 | BO | WS | 1 | 0 |
| 0.75 | ML | UL | 1 | 1 |
| 0.75 | AT | WS | 0 | 1 |
| 0.76 | BO | ML | 1 | 0 |
| 0.76 | MG | BO | 0 | 1 |
| 0.77 | BO | AT | 1 | 0 |
| 0.77 | BO | SW | 1 | 0 |
| 0.79 | BO | UL | 1 | 1 |
| 1 | MG | WS | 1 | 1 |
| 1 | BO | SWRS | 1 | 1 |
| **Superficial BCC** | | | | |
| 1 | ML | SW | 1 | 1 |

**FIGURE LEGENDS**

**Figure I. Probability distributions highlighting the relationship between BCC subtypes and dermoscopic patterns.**

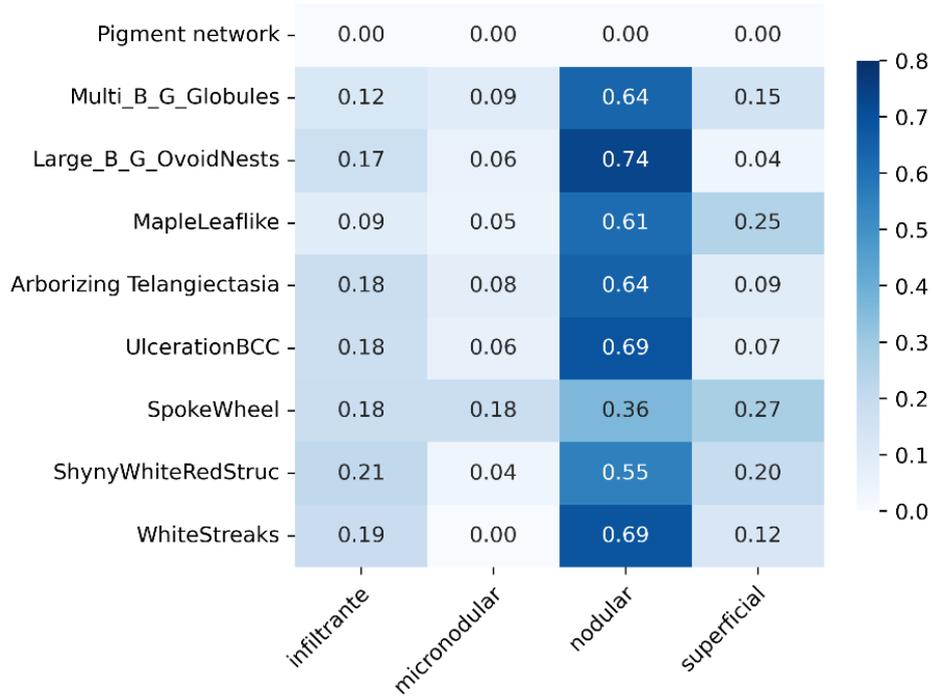

**Figure II present $P(\text{class} \mid \text{pattern})$ for BCC subtypes. The nodular subtype consistently shows high probabilities (>0.6) across most patterns, excluding SW and SWRS, making it the most distinguishable subtype in this analysis.**

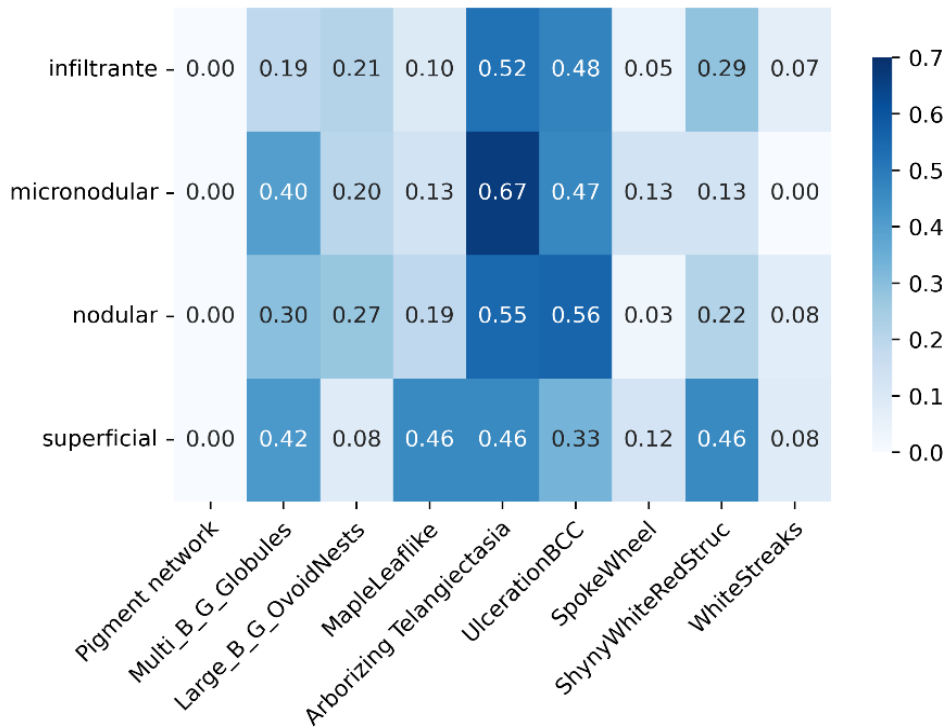

**Figure I.II present $P(pattern \mid \text{class})$ for dermoscopic patterns across BCC subtypes. Patterns AT and UL are prevalent in all subtypes, serving as general indicators of BCC but providing limited specificity for subtype differentiation.**

**Figure II.** Co-occurrence matrix representing the frequency of dermoscopic patterns observed for each BCC subtype (Infiltrative, Micronodular, Nodular, and Superficial), highlighting the unique pattern distribution associated with each subtype.

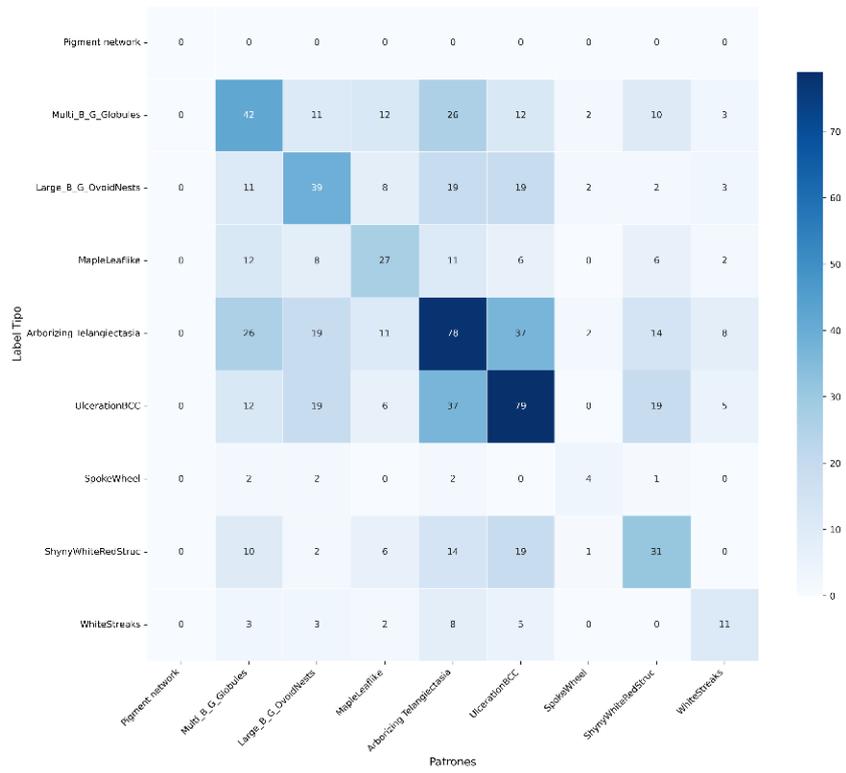
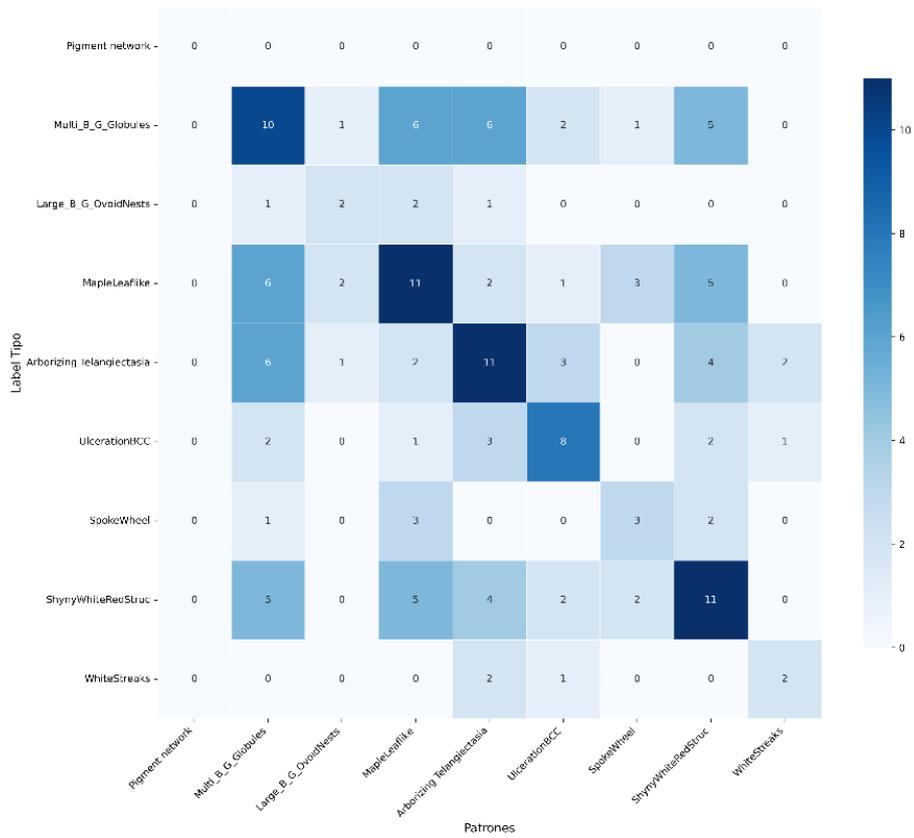